\documentclass[%
 reprint,
 amsmath,amssymb,
 aps,
prb,
]{revtex4-1}

\usepackage{graphicx}% Include figure files
\usepackage{dcolumn}% Align table columns on decimal point
\usepackage{bm}% bold math
\usepackage{amsmath}
\usepackage{color}
\begin{document}

\preprint{APS/123-QED}

\title{Degenerate ground state in the classical pyrochlore antiferromagnet Na$_3$Mn(CO$_3$)$_2$Cl}% Force line breaks with \\
\author{Kazuhiro~Nawa$^{1}$}
\email{knawa@tohoku.ac.jp}
\author{Daisuke~Okuyama$^{1}$}
\author{Maxim~Avdeev$^{2,3}$}
\author{Hiroyuki~Nojiri$^{4}$}
\author{Masahiro~Yoshida$^{5}$}
\author{Daichi Ueta$^{5}$}
\author{Hideki Yoshizawa$^{5}$}
\author{Taku J. Sato$^{1}$}
\affiliation{%
$^{1}$Institute of Multidisciplinary Research for Advanced Materials, Tohoku University, 2-1-1 Katahira, Sendai 980-8577, Japan \\
$^{2}$Australian Nuclear Science and Technology Organisation, Kirrawee DC, NSW 2232, Australia \\
$^{3}$School of Chemistry, The University of Sydney, Sydney 2006, Australia \\
$^{4}$Institute for Materials Research, Tohoku University, 2-1-1 Katahira, Sendai 980-8577, Japan \\
$^{5}$Institute for Solid State Physics, The University of Tokyo, Kashiwa, Chiba 277-8581, Japan
}%

\date{\today}% It is always \today, today,

\begin{abstract}
In an ideal classical pyrochlore antiferromagnet without perturbations, an infinite degeneracy at a ground state leads to absence of a magnetic order and spin-glass transition.
Here we present Na$_3$Mn(CO$_3$)$_2$Cl as a new candidate \textcolor{black}{compound where classical spins are coupled antiferromagnetically on the pyrochlore lattice}, and report its structural and magnetic properties.
The temperature dependences of the magnetic susceptibility and heat capacity, and the magnetization curve are consistent with those of an $S$ =~5/2 pyrochlore lattice antiferromagnet with nearest-neighbor interactions of 2~K.
Neither an apparent signature of a spin-glass transition nor a magnetic order is detected in magnetization and heat capacity measurements, or powder neutron diffraction experiments.
On the other hand, an antiferromagnetic short-range order from the nearest neighbors is evidenced by the $Q$-dependence of the diffuse scattering which develops around 0.85~\AA$^{-1}$.
A high degeneracy near the ground state in Na$_3$Mn(CO$_3$)$_2$Cl is supported by the magnetic entropy estimated as almost 4 J K$^{-2}$ mol$^{-1}$ at 0.5 K.
\end{abstract}

\pacs{Valid PACS appear here}% PACS, the Physics and Astronomy
\maketitle

%\tableofcontents

\section{\label{introduction}Introduction}
Frustrated magnets have been receiving attention for decades since competing interactions destabilize conventional magnetic order and instead induce non trivial magnetic order,
quantum phases, and critical phenomena\cite{frustration1, frustration2, frustration3, frustration4, frustration5}.
A classical pyrochlore magnet belongs to a class of the system that exhibits interesting magnetic properties\cite{Py0, Py1, Py2, Py3, Py4, Py5, allinallout}.
For the pyrochlore magnet where \textit{Ising} spins are coupled by \textit{ferromagnetic} nearest-neighbor interactions, 
two-in/two-out ground states on a single tetrahedron lead to a highly degenerate ground state as a whole\cite{allinallout}, which is called a spin-ice\cite{spinice1, spinice2, spinice3}.
%In addition, dipolar interaction in this system can be fractionalized into magnetic monopoles\cite{spinice4}.
In case of \textit{Heisenberg} spins with \textit{antiferromagnetic} nearest-neighbor interactions, 
the ground state is also highly degenerate, and, moreover, ground state manifolds can be transformed into each other by rotating spins without energy change\cite{Py3, Py4}.
Thus, ideally, a classical Heisenberg pyrochlore antiferromagnet exhibits neither a long-range magnetic order nor spin-glass transition as $T \to~0$.
However, such a large degeneracy can be easily lifted by small perturbations.
For instance, a spin-glass state is induced by bond disorder between magnetic ions\cite{spinglass1, spinglass2, spinglass3},
a nematic state can occur from bi-quadratic interactions\cite{nematic},
and a N\'eel order, partial order, and multi-$Q$ magnetic orders from next-nearest neighbor or third-nearest neighbor interactions \cite{PyJ1J21, PyJ1J22, PyJ1J23, PyJ1J24}.
In real materials, these weak interactions often result in magnetic order or spin-glass state.
%, as in a representative pyrochlore antiferromagnet FeF$_3$.
%In this compound, a short range order develops below 160~K\cite{FeF3_2} and an all-in-all-out magnetic order exhibits at $T_\mathrm{N}$ = 15--22~K\cite{FeF3_1, FeF3_3}.
%The reason why the all-in-all-out magnetic order is stabilized is not clear: a weak single-ion anisotropy may be present\cite{FeF3_2} or both Dzyaloshinskii Moriya and bi-quadratic interactions can be its origin\cite{FeF3_4}.
For instance, the representative pyrochlore antiferromagnet FeF$_3$\cite{FeF3_2} exhibits an all-in-all-out magnetic order at $T_\mathrm{N}$ = 15--22~K\cite{FeF3_1, FeF3_3} possibly due to a weak single-ion anisotropy\cite{FeF3_2}
or coexisting Dzyaloshinskii Moriya and bi-quadratic interactions\cite{FeF3_4}.
Some magnetic orders or spin-glass states are also found in \textcolor{black}{other antiferromagnets} with highly Heisenberg interactions, such as
Mn$_2$Sb$_2$O$_7$\cite{FeF3_2, Mn2Sb2O7_1, Mn2Sb2O7_2}, Gd$_2$Ti$_2$O$_7$\cite{Gd2Ti2O7_1, Gd2Ti2O7_2, Gd2Ti2O7_3}, ZnFe$_2$O$_4$\cite{ZnFe2O4_1, ZnFe2O4_2}, and NaSrMn$_2$F$_7$\cite{F_py}.

\begin{figure}[t]
\includegraphics[width=8cm]{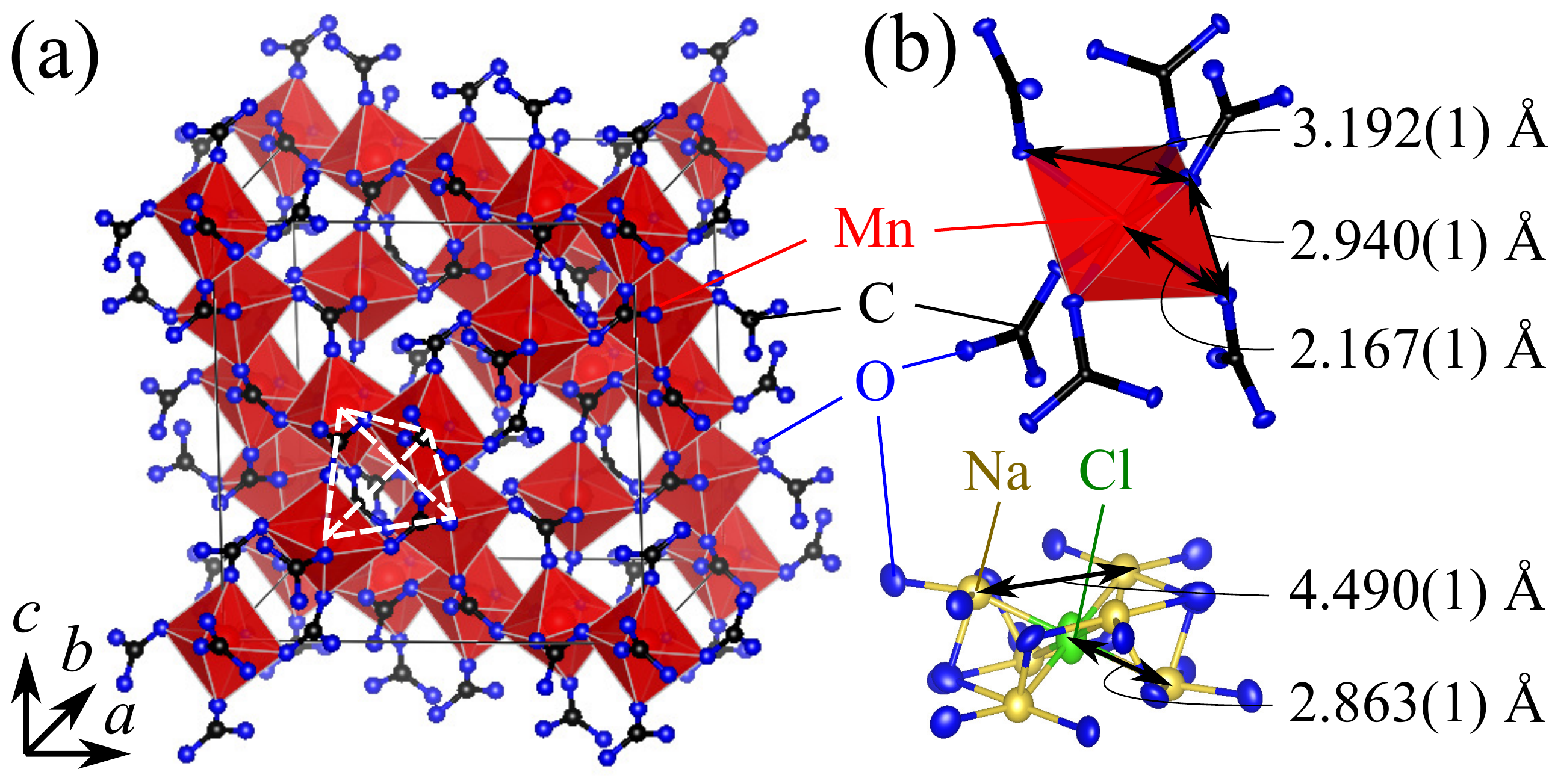}
\caption{\label{cryst} Crystal structure of Na$_3$Mn(CO$_3$)$_2$Cl. Mn, C, O, Na, and Cl atoms are shown by red, black, blue, dark yellow and green ellipsoids, reflecting atomic displacement parameters of each atom.
(a) Arrangement of MnO$_6$ octahedra in a unit cell. MnO$_6$ octahedra are connected by carbonate ions, forming a pyrochlore network of Mn atoms. 
\textcolor{black}{One of a tetrahedron forming the pyrochlore lattice is indicated by white dashed lines.}
Na and Cl atoms are omitted for clarity.
(b) Local structures around Mn and Cl atoms.
VESTA program is used for visualization\cite{VESTA}.}
\end{figure}

In this paper, we report structural and magnetic properties of \textcolor{black}{Na$_3$Mn(CO$_3$)$_2$Cl,
a new Heisenberg antiferromagnet where classical spins are placed on the pyrochlore lattice.}
The crystal structure of Na$_3$Mn(CO$_3$)$_2$Cl is shown in Fig.~\ref{cryst}.
It is a Mn-analog of Na$_3$Co(CO$_3$)$_2$Cl:
\textcolor{black}{
a pyrochlore network of magnetic ions is formed by $T$O$_6$ octahedra linked by carbonate ions ($\mathrm{CO_3^{2-}}$)\cite{PyCo1}.
Although its crystal structure is not exactly identical with that of a pyrochlore structure
(an ordered derivative of the fluorite structure), 
its magnetism should be equivalent to that of a pyrochlore antiferromagnet.
From this reason, we simply call both pyrochlore antiferromagnets in the following.}
For Na$_3$Co(CO$_3$)$_2$Cl, several phase transitions are detected through macroscopic measurements and powder neutron scattering experiments\cite{PyCo2}.
In particular, an all-in-all-out magnetic order occurs as temperature is decreased below $T_\mathrm{N}$ =~1.5~K.
Coexistence of antiferromagnetic nearest-neighbor interactions $J$ and weak ferromagnetic next-nearest-neighbor interactions $J'$ may lead to the successive magnetic transitions\cite{PyCo2}.
In addition, orbital degeneracy should be also a key since a coupling between the orbital degeneracy and a spin 3/2 for Co$^{2+}$ (3$d^7$) selects a Kramers doublet as a ground state.
%A temperature dependence of an ac-magnetic susceptibility exhibits a broad peak at $T_\mathrm{a}$ =~4.5~K.
%The peak position is frequency dependent, which is characteristic of a spin-glass transition.
%As a temperature is decreased below $T_\mathrm{N}$ =~1.5~K, magnetic Bragg peaks appear at $\mathbf{q}$ = $\mathbf{0}$, in addition to the diffuse scattering which develops below 30~K.
%The symmetry analysis indicates occurrence of the all-in-all-out long-range magnetic order.
%Although a fit to the diffuse scattering yields dominant antiferromagnetic nearest-neighbor interactions $J$ and weak ferromagnetic next-nearest neighbor interactions $J'$\cite{PyCo2}, 
%ferromagnetic $J'$ should not lead to the all-in-all-out magnetic order but instead stabilizes the multi-$Q$ magnetic order with a magnetic wave vector $\mathbf{q}$ = ($h$ $h$ 0) ($h~\sim$ 3/4)\cite{PyJ1J23, PyJ1J24}.
On the other hand, Na$_3$Mn(CO$_3$)$_2$Cl consists of Mn$^{2+}$ (3$d^5$), which should have little anisotropy because of the absence of the orbital degeneracy.
Thus, this compound is appropriate to investigate magnetic properties of a classical pyrochlore antiferromagnet experimentally.
To examine the ground state of Na$_3$Mn(CO$_3$)$_2$Cl, we performed magnetization and heat-capacity measurements, and powder neutron diffraction experiments.
We present these results in the following sections,
and discuss magnetic properties of Na$_3$Mn(CO$_3$)$_2$Cl as a classical pyrochlore antiferromagnet with a large degeneracy near the ground state.
%In spite of the development of antiferromagnetic correlations below 50~K,
%neither the signature of a spin-glass nor a magnetic order was not detected above 2~K and 0.05~K, respectively, in contrast to Na$_3$Co(CO$_3$)$_2$Cl.
%In addition, an estimation of the magnetic entropy indicates a large degeneracy near the ground state.
%In this sense, we propose that Na$_3$Mn(CO$_3$)$_2$Cl can exhibit an interesting ground state such as a spin liquid state expected in a classical pyrochlore antiferromagnet.

\section{\label{experiments}Experiments}
A Na$_3$Mn(CO$_3$)$_2$Cl sample was synthesized by a solvothermal method.
First, 2~mmol of Na$_2$CO$_3$ (212.0~mg) and 1~mmol of NaHCO$_3$ (84.0~mg) were dissolved in pure water of 0.56~ml, and
then 1~mmol of MnCl$_2\cdot4$H$_2$O (197.9~mg) was added and the mixture was stirred until the solution became homogeneous.
Next, 9~ml of ethanol and 2~mmol of 1,10-phenanthroline monohydrate (317.2~mg) were added, with 10 min stirring, after each procedure.
The mixed solution was poured into a 25~ml-volume Teflon container together with a magnetic stirrer, sealed in a stainless vessel, and heated at 160~$^\circ$C for 4 days with stirring the solution with a speed of 150~rpm.
After the heat treatment, the solution was filtered, and the sample was further washed several times to remove extra carbonates, salts and organic materials.
For this purpose, both ethylene glycol and the mixed solution of pure water (2.24~ml) and ethanol (36~ml) were selected as solvents.
An aggregate of brown octahedral crystals was left after the whole procedure. 

The crystal structure was determined from a single-crystal synchrotron X-ray diffraction (XRD) experiment performed at BL-8A, Photon Factry KEK using a 30~$\mu$m-length octahedral crystal.
An energy of an incident beam was tuned to 12.4~keV ($\lambda =~0.9999$~\AA ) and scattered X-rays were collected with a cyllinder-shaped imaging plate.
Nitrogen gas was flowed toward the crystal to avoid absorbing moisture and heated to keep the temperature to 300~K.
An $|F|$ table down to a resolution limit of 0.5~\AA \ was made through RAPID-offline program (Rigaku), using anomalous scattering factors calculated relativistically\cite{AnomXRD}.
The crystal structure was determined by a direct method using SIR2004 program\cite{SIR} and refined using Crystals program.
% using CRYSTALSTRUCTURE program (Rigaku).

Physical properties were measured by using several batches of the polycrystalline sample synthesized by the solvothermal method.
Both dc- and ac-magnetic susceptibilities were measured by Magnetic Properties Measurement System (Quantum Design), and
heat capacity was measured by a relaxation method using Physical Properties Measurement System (Quantum Design).
The low-temperature probe equipped with an adiabatic-demagnetization refrigerator was used to measure the magnetic susceptibility below 2~K\cite{LTprobe}.
High-field magnetization curve was measured using a pulse magnet combined with a $^3$He cryostat to reach down to 0.5~K.
To investigate whether a magnetic order is present or not at low temperatures microscopically, powder neutron diffraction measurements were performed at high-resolution powder diffraction spectrometer ECHIDNA\cite{ECHIDNA}.
Neutrons with wavelength $\lambda$ = 2.4395 \AA \ were selected by a monochromator using Ge 331 reflections.
Neutron diffraction patterns at 0.05 and 2~K were obtained by using a dilution refrigerator.
To improve thermal conductivity, deuterated isopropanol ($d$-isopropanol) was put in a copper can together with 0.9~g of polycrystalline sample.
%0.9~g of polycrystalline samples are put inside a copper can together with deuterated isopropanol ($d$-isopropanol) to improve thermal conductivity,
In addition, neutron diffraction patterns were collected at 1.5--200~K from 1.3~g of polycrystalline sample set in a vanadium can, using a top-loading cryostat.
Rietveld analysis was made by Fullprof suite\cite{Fullprof}.

\section{\label{CS}Crystal structure}
\begin{table}[b]
\begin{center}
\caption{\label{Refinement}Refinement parameters for Na$_3$Mn(CO$_3$)$_2$Cl. A full-matrix least-squares method on $|F|^2$ is adopted for the refinement.}
\begin{tabular}{lc} \hline\hline
2$\theta$ range for data collection & 7.0--144.5$^\circ$ \\
Index ranges & -24 $\leq h \leq$ 23\\
& -27 $\leq k \leq$ 27 \\
& -21 $\leq l \leq$ 22 \\
Reflections collected & 8766 \\
Independent reflections & 855 \\
Completeness to $\theta$ & 0.868\footnote{\textcolor{black}{The completeness is lower than standard since a spindle axis is fixed and thus a single-axis rotation is only possible.}} \\
Data/restraints/parameters & 706/0/23 \\
Goodness of fit on $|F|^2$ & 0.956 \\
$R$ indices for all data ($R$, $R_w$) & 0.0136, 0.0247 \\
Extinction coefficient & 6.1(10) \\
Largest diff. peak and hole (e \AA$^{-3}$) & 0.35 and -0.39 \\
\hline\hline
\end{tabular}
\end{center}
\end{table}

\begin{table*}[t]
\caption{\label{Structure}Structural parameters of Na$_3$Mn(CO$_3$)$_2$Cl at 300~K determined from the single-crystal XRD experiment.
The space group is $Fd\overline{3}$ (No.~203), and the lattice parameter is $a$ = 14.1932(3)~\AA .
Atomic coordinations are represented by fractional coordinates.
Anisotropic displacement parameters $U_{11}$, $U_{22}$, $U_{33}$, $U_{12}$, $U_{13}$, and $U_{23}$,
and equivalent isotropic displacement parameters $B_\mathrm{eq}$ [$B_\mathrm{eq}$ = 8$\pi^2$/3 $\times$ ($U_{11}+U_{22}+U_{33}$)]
are listed in units of \AA $^2$. Occupancy is fixed to 1 for all atoms.}
\label{atom}
\begin{center}
\begin{tabular}{lcccccccccccc}
\hline\hline
 & site & $x$ & $y$ & $z$ & $U_{11}$ & $U_{22}$ & $U_{33}$ & $U_{12}$ & $U_{13}$ & $U_{23}$ & $B_\mathrm{eq}$ \\
\hline
Na & 48$f$ & 1/8 & 0.34864(2) & 5/8 & 0.01663(10) & 0.01788(11) & 0.01907(11) & 0 & -0.00337(8) & 0 & 1.410(4) \\
Mn & 16$c$ & 0 & 1/2 & 1/2 & 0.00952(3) & 0.00952(3) & 0.00952(3) & -0.00043(2) & -0.00043(2) & -0.00043(2) & 0.752(1) \\
C & 32$e$ &0.03054(2) & 0.71946(2) & 0.53054(2) & 0.00864(7) & 0.00864(7) & 0.00864(7) & -0.00043(7) & 0.00043(7) & -0.00043(7) & 0.682(2) \\
O & 96$g$ &-0.02533(2) & 0.65000(2) & 0.51505(2)	& 0.01300(10) & 0.01066(9) & 0.01845(11) & -0.00377(7) & 0.00011(8) & -0.00237(8) & 1.108(3) \\
Cl & 16$d$ & 1/4	 & 1/4 & 1/2 & 0.02573(7) & 0.02573(7) & 0.02573(7) & 0.01009(7) & -0.01009(7) & -0.01009(7) & 2.032(2) \\
\hline\hline
\end{tabular}
\end{center}
\end{table*}

\begin{figure}[t]
\includegraphics[width=8cm]{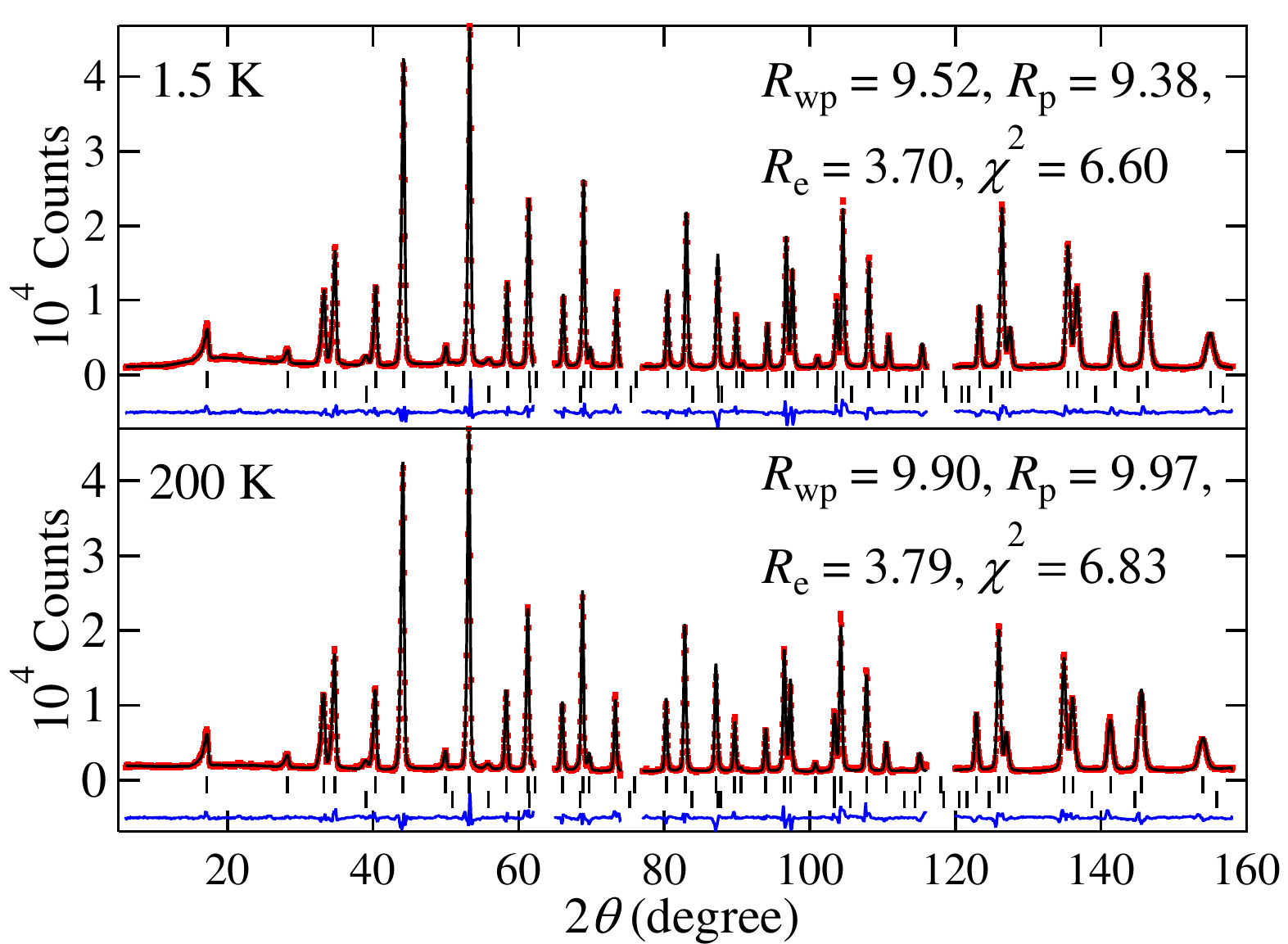}
\caption{\label{NPD} Powder neutron diffraction pattern at 1.5 and 200~K collected by ECHIDNA together with Rietveld analysis.
Total neutron counts are normalized by $1.5 \times~10^6$ ncu.
Observed, calculated, and the difference between both intensities are shown by red points, black, and blue curves, respectively. 
Reflection position for Na$_3$Mn(CO$_3$)$_2$Cl and MnCO$_3$ expected from a wavelength of 2.4395 \AA \  are indicated by the upper and lower vertical lines, respectively.
Reflections from Al cryostats are removed in the figure.}
\end{figure}

Refinement and structural parameters determined from the single-crystal XRD analysis are summarized in Table \ref{Refinement} and \ref{Structure}, respectively.
The crystal structure of Na$_3$Mn(CO$_3$)$_2$Cl is shown in Fig.~\ref{cryst}.
Structural features of Na$_3$Mn(CO$_3$)$_2$Cl are the same as those of Na$_3$Co(CO$_3$)$_2$Cl\cite{PyCo1, PyCo2}: 
$T$O$_6$ ($T$ =~ Co, Mn) octahedra are connected by carbonate ions to form the pyrochlore lattice.
In addition, local structures around Mn(Co) and Cl atoms are also quite similar.
Mn atoms coordinate with six carbonate ions, forming MnO$_6$ octahedra compressed trigonally along [111] or its equivalent directions.
A length of Mn--O bond is 2.170~\AA \ and quite close to 2.190~\AA \ in MnCO$_3$ \cite{MnCO3}, where Mn atoms are also linked with six carbonate ions.
Cl atoms are characterized by a large and anisotropic atomic displacement parameter: 
in particular, it becomes large along [111] or its equivalent directions, while it becomes small in the perpendicular direction.
This feature indicates specific coordination around Cl atoms rather than large deficiency of Cl atoms themselves.
There are 6 Na atoms around Cl atoms with a bond length of 2.873~\AA, which is a little larger than 2.820~\AA \ in NaCl\cite{NaCl}.
In addition, large distortion of Na$_6$Cl octahedra yields a large vacancy on [111] or its equivalent directions, resulting in mobile Cl atoms.
Consistency between single-crystal XRD and powder neutron diffraction experiments was confirmed
by applying Rietveld refinement method on the powder neutron diffraction patterns.
Starting from initial parameters based on the crystal structure determined from the single crystal XRD analysis, 
powder neutron diffraction patterns are well reproduced with very little change onf atomic positions (Fig.~\ref{NPD}).
The refined lattice constants are $a$ = 14.1353(1)~\AA \ and 14.1631(3)~\AA \ at 1.5~K and 200~K, respectively.
Several very weak peaks due to a $\lambda/2$ component are observed, which are included in the refinement.
In addition, 1.8 wt\% of MnCO$_3$ is detected as an impurity phase.
As the temperature is decreased from 200~K to 1.5~K, a broad peak around 20$^\circ$ develops at low temperatures, indicating enhancement of antiferromagnetic correlations. 
%Although no strong signature of magnetic Bragg peaks were found.
The detail is discussed in section~\ref{NPDexp}.

\section{\label{MTP}Magnetic and thermodynamic properties}
\begin{figure}[t]
\includegraphics[width=8cm]{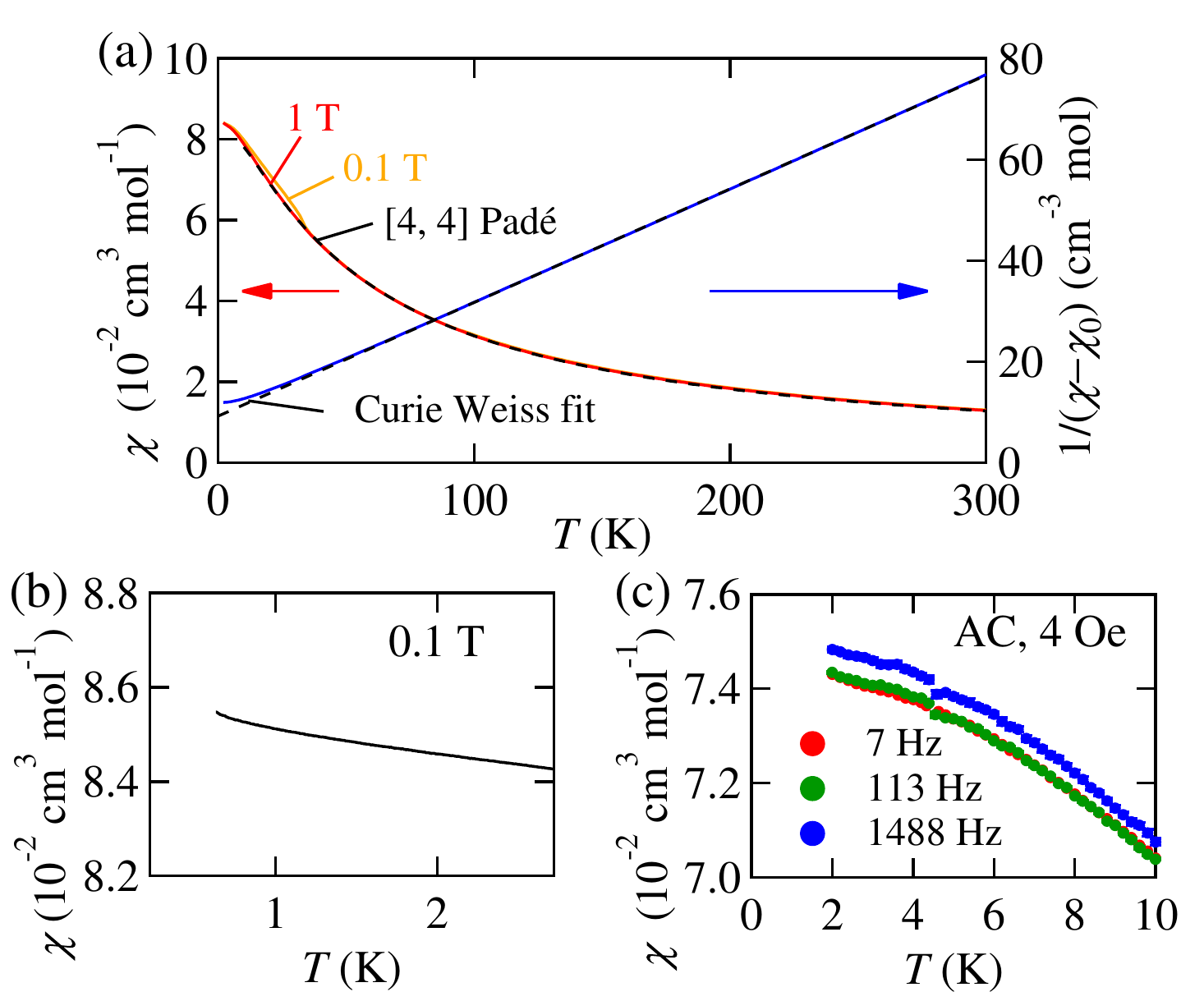}
\caption{\label{chi} (a) Temperature dependence of the magnetic susceptibility measured at 1~T (red) and 0.1~T (orange) with zero-field cooling.
The blue curve indicates the inverse magnetic susceptibility. Dashed and dotted curves represent a fit to the [4, 4] Pad\'e approximant for a high-temperature series expansion of a pyrochlore lattice\cite{HTSE}, and a Curie-Weiss fit, respectively.
(b) Magnetic susceptibility down to 0.6~K measured by the low-temperature probe after cooling in the field of 5~T and then decreasing the field to 0.1~T. (c) AC susceptibility below 10~K.}
\end{figure}

A temperature dependence of the magnetic susceptibility is well explained by that of a $S$ = 5/2 pyrochlore antiferromagnet,
as shown in Fig.~\ref{chi}(a).
At low fields, magnetic susceptibility exhibits an anomaly at 34~K, which is due to a ferrimagnetic transition of MnCO$_3$
included as the impurity phase\cite{MnCO3_2}.
To reduce the effect of the impurity phase, the following analysis is performed on the data measured at 1~T.
As shown by a blue solid line in Fig.~\ref{chi}(a), the inverse susceptibility above 50~K well follows the Curie-Weiss law.
From a fit of the magnetic susceptibility at 50--300~K to the Curie-Weiss law,
the effective moment and a Weiss temperature are estimated as $\mu^2_\mathrm{eff}$ = 35.6(1), and $\Theta_\mathrm{W}$ =~40.9(1)~K, respectively.
The temperature independent susceptibility is fixed to $-$1.23 $\times$ 10$^{-4}$ cm$^3$ mol$^{-1}$ from Pascal constants of the constituent atoms.
The effective moment is consistent with 35 expected for $S$ =~5/2 and an isotropic $g$ value of 2.
\textcolor{black}{Below 50~K, the magnetic susceptibility becomes slightly lower than that expected from the Curie-Weiss law, which is consistent with that expected for a classical pyrochlore antiferromagnet.
The similar temperature dependence is also observed in Gd$_2$Ti$_2$O$_7$\cite{Gd2Ti2O7_1},
while the magnetic susceptibility becomes larger than that expected from the Curie-Weiss law in Mn$_2$Sb$_2$O$_7$\cite{Mn2Sb2O7_1, Mn2Sb2O7_2} and NaSrMn$_2$O$_7$\cite{F_py}.}
The data at 10--300~K are well fitted by a [4, 4]-Pad\'e approximant applied to high-temperature-series expansion of $S$ =~5/2 pyrochlore antiferromagnet\cite{HTSE}. 
The fit yields a nearest-neighbor interaction $J$ and a g-value of $J$ = 2.03(1)~K and $g$ = 1.99(1), respectively.

At low temperatures, signatures of a magnetic order and spin-glass transition are not observed, as presented by Figs.~\ref{chi}(b) and \ref{chi}(c). 
Figure~\ref{chi}(b) shows the magnetic susceptibility measured at 0.6--3~K using the adiabatic-demagnetization refrigerator.
As the temperature decreases, the magnetic susceptibility slightly increases but does not show any anomaly which would suggest a magnetic transition.
Figure~\ref{chi}(c) shows the ac magnetic susceptibility measured at 2--10~K with frequencies of 7--1488~Hz.
The jump observed at 4.5~K is not from the sample but due to small instability of the temperature around the boiling point of helium.
Spin-glass-like features such as occurrence of a broad peak and a frequency dependence of its peak position are not observed
in this frequency range.

\begin{figure}[t]
\includegraphics[width=7cm]{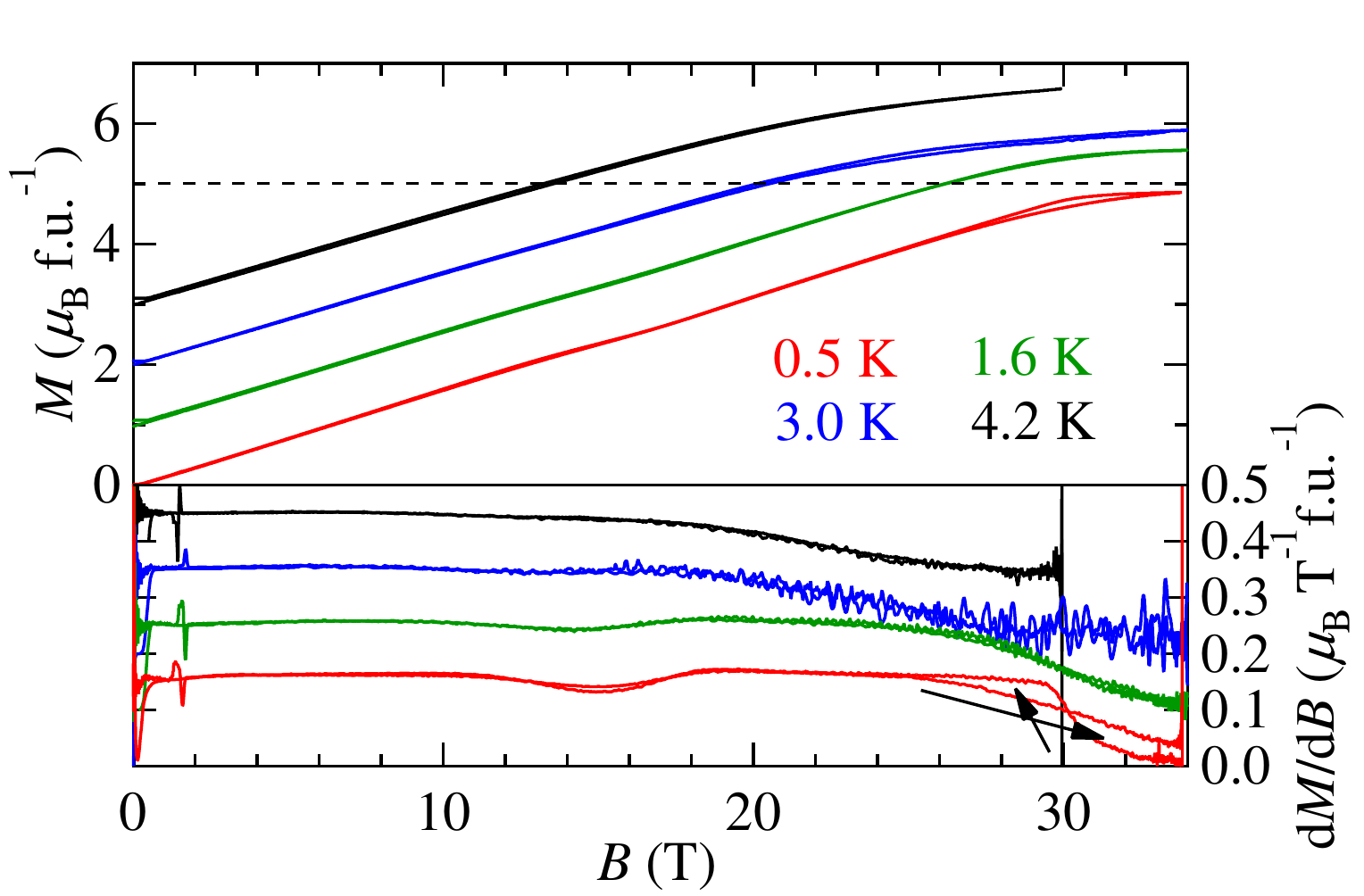}
\caption{\label{MH} Magnetization curves (upper) and its field derivative (lower) measured at 0.5 (red), 1.6 (green), 3.0 (blue), and 4.2~K (black). The curves measured at higher temperatures are shifted for clarity.
Arrows indicate a time sequence of magnetization process around a hysteresis loop.}
\end{figure}

High-field magnetization curves also provide $J$ consistent with that estimated from the magnetic susceptibility.
Figure~\ref{MH} shows high-field magnetization curves measured by a pulse field.
At 0.5~K, the magnetization linearly increases with an increasing  magnetic field as expected from a mean-field approximation:
the sum of four spins on each tetrahedron, and thus, the total magnetization should be proportional to the magnetic field. 
The magnetization curve saturates at the saturation field $B_\mathrm{s}$ of 30.6(3)~T.
According to a mean-field approximation, a saturation field is related with $J$ as $g \mu_\mathrm{B} B_\mathrm{s}$ = $8 J S$, where $\mu_\mathrm{B}$ is a Bohr magnetron.
Thus, from $B_\mathrm{s}$ =~30.6~T and $g$ =~1.99, $J$ can be roughly estimated as 2.0(1)~K, which is consistent with that estimated from the magnetic susceptibility.
A hysteresis is present in the magnetization curve, which is due to a magnetocaloric effect: the temperature is increased (a decreased) with an increasing (decreasing) magnetic field  under a quasi-adiabatic condition because of remaining degeneracy below the magnetic saturation.
The slope of the magnetization curve near the magnetization saturation becomes smaller at higher temperatures, indicating smaller degeneracy at high fields. 
In addition, the slope of the magnetization curve slightly decreases at half-magnetization.
The decrease becomes prominent at low temperatures, suggesting the anomaly is not caused by an order-by-disorder mechanism.
Instead, we speculate that small anisotropic interactions or weak spin-lattice couplings may be present.
It is noted that spin-lattice couplings can stabilize three-up/one-down configurations for each tetrahedron under a magnetic field, leading to 1/2 plateau in a magnetization curve\cite{spinlattice, spinlattice2}.
For the magnetization curve of Na$_3$Mn(CO$_3$)$_2$Cl, the field range where the decrease of the slope is observed is limited to a width of about 3.5~T.
This means that biquadratic interactions as small as 3\% of $J$ can be present if the spin-lattice couplings cause the anomaly in magnetization.
Biquadratic interactions never stabilize a magnetic order\cite{nematic}, and thus its presence does not contradict with the following discussion.

\begin{figure}[t]
\includegraphics[width=8cm]{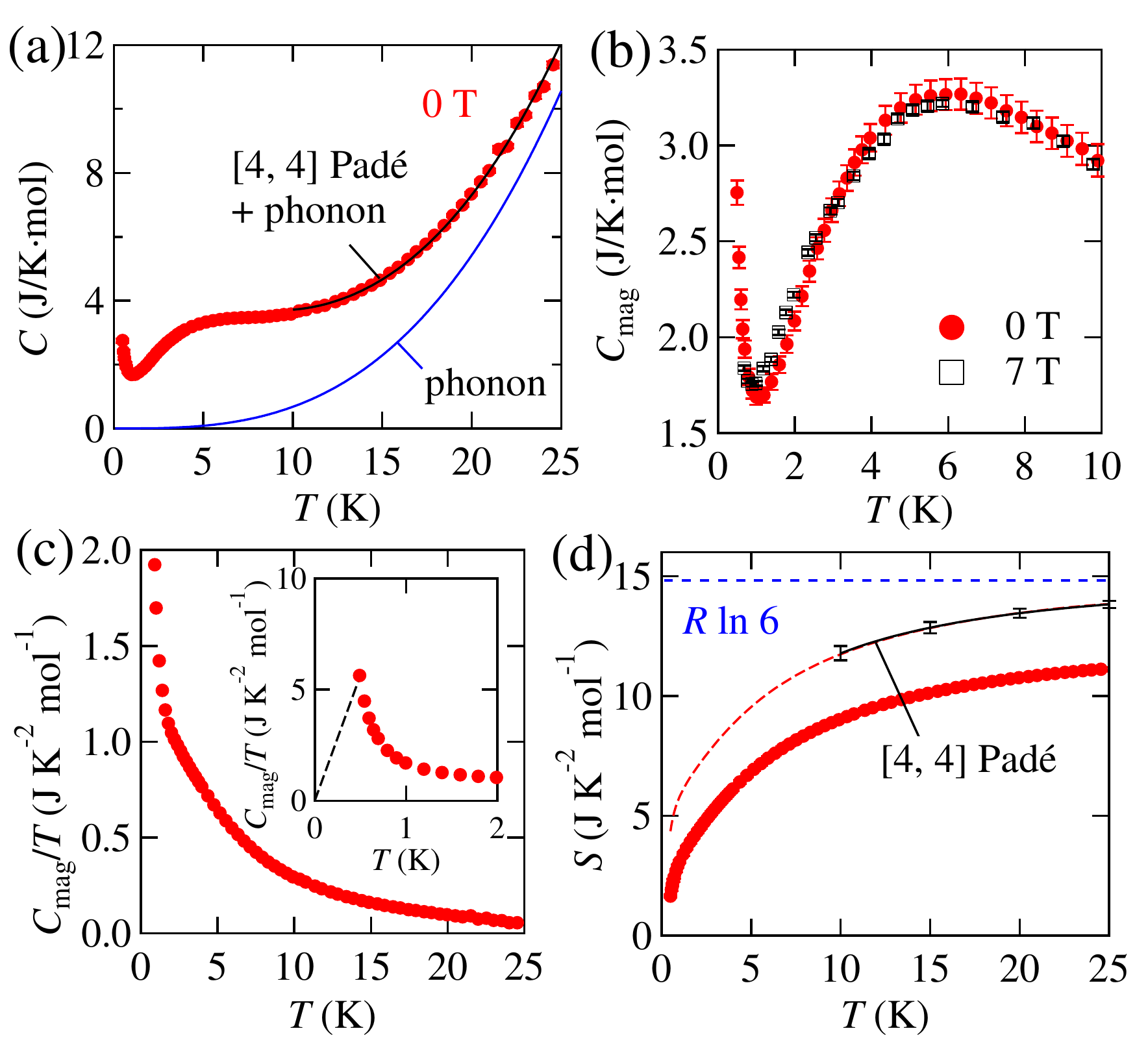}
\caption{\label{HC} \textcolor{black}{(a) Temperature dependence of heat capacity at 0~T. A black solid curve indicates the fit including both phonon and magnetic contributions,
while a blue solid curve represents the phonon contribution estimated from the fit (see main texts for details).
%the fit as a sum of heat capacity of a pyrochlore lattice\cite{HTSE} with $J = 1.9$~K, which is obtained from a fit 
(b) Magnetic heat capacity at 0~T (solid circles) compared with that of 7~T (open squares).}
(c) Magnetic heat capacity divided by the temperature $C_\mathrm{mag}/T$ at 0~T.
The inset shows an enlarged view below 2~K.
A black dashed line corresponds to the extension of $C_\mathrm{mag}/T$ below 0.5~K just for the estimation of the magnetic entropy.
(d) The magnetic entropy determined from integrating $C_\mathrm{mag}/T$.
A black dashed line in the inset of Fig.~\ref{HC}(c) leads to the estimate indicated by red solid circles.
Red dashed curves are obtained by shifting the estimated entropy above so that they match the black solid curve which indicates the magnetic entropy expected for a pyrochlore lattice with $J = 2.0$~K.
The error bar of the black solid curve corresponds to entropy difference from 0.2~K difference in $J$.}
%estimated by subtracting the high-temperature contribution from a total entropy, $R$ ln 6.}
\end{figure}

%The magnetic susceptibility and the magnetization curve reveal that magnetic properties of Na$_3$Mn(CO$_3$)$_2$Cl are well described by a $S$ =~5/2 pyrochlore antiferromagnet.
%Since signatures of a magnetic transition and spin glass are not observed, a large degeneracy should be left near the ground state,
%and in fact, it is evidenced by heat-capacity measurements.
The temperature dependence of a heat capacity %not only supports the absence of the magnetic order and but also
indicates a large degeneracy near the ground state.
\textcolor{black}{A temperature dependence of a heat capacity $C$ is shown in Fig.~\ref{HC}(a).
The heat capacity exhibits a broad peak at low temperatures, as is usual for pyrochlore antiferromagnets
\cite{Mn2Sb2O7_2, Gd2Ti2O7_1, F_py, PyCo2}.
As shown in Fig.~\ref{HC}(b), the peak is slightly broadened in 7~T, indicating that frustration is partially relieved by the magnetic field.}
The total heat capacity measured at 10--25~K is fitted with the sum of phonon and magnetic contributions: 
the former and the latter are estimated by Debye model and a [4, 4] Pad\'e approximant of the high-temperature series expansion\cite{HTSE}, respectively.
\textcolor{black}{The fit is represented by the black curve in Fig.~\ref{HC}(a); note that it is shown only above 10~K since the approximant is not reliable below the temperature of $J S (S+1)/2$.
The heat capacity at 10--25~K agrees well with the fitted curve.}
This fit yields Debye temperature $\Theta_\mathrm{D}$ of \textcolor{black}{334(5)~K} and nearest-neighbor interactions $J$ of \textcolor{black}{2.0(1)~K}, which are consistent with magnetization measurements.

The magnetic contribution is extracted by subtracting the phonon contribution (blue curve in Fig.~\ref{HC}(a)), resulting in 
a temperature dependence of a magnetic heat capacity divided by temperature, $C_\mathrm{mag}/T$, which is shown as red circles in Fig.~\ref{HC}(c).
%Phonon contributions are estimated by fitting a heat capacity measured at 10--25~K with the sum of the 
%Debye model with Debye temperature of $\Theta_\mathrm{D}$ =~380 $\pm$ 20~K,
%which is referred from that of Na$_3$Co(CO$_3$)$_2$Cl\cite{PyCo2}. 
%The uncertainty of the phonon contributions are also included as error bars of $C_\mathrm{mag}/T$ in Fig.~\ref{HC}(a).
%A black dashed curve represents $C_\mathrm{mag}/T$ of a $S$ =~5/2 pyrochlore antiferromagnet with $J$ = 2.0 $\pm$ 0.2,
%which is simulated from a [4, 4] Pad\'e approximant of the high-temperature series expansion\cite{HTSE}.
As temperature is further decreased, $C_\mathrm{mag}/T$ increases gradually down to 2~K.
Below 2~K, $C_\mathrm{mag}/T$ increases greatly but never exhibits a peak which would indicate a magnetic transition, as shown in an inset of Fig.~\ref{HC}(c).
In fact, this is consistent with the neutron diffraction experiments which did not show an apparent signature of the magnetic order even at 0.05~K, as will be discussed later.
Since the temperature below 0.5~K was not accessible in the present experiment,
$C_\mathrm{mag}/T$ is assumed to decrease linearly down to 0 as a black dashed curve in the inset of Fig.~\ref{HC}(c).
%the following temperature dependences are assumed to estimate a magnetic entropy: $C_\mathrm{mag}/T$ at 0~K becomes equal to 
%(1) 0, (2) $C_\mathrm{mag}/T$ (0.5~K), and (3) double of $C_\mathrm{mag}/T$ (0.5~K), and $C_\mathrm{mag}/T$ depends on $T$ linearly between 0 and 0.5~K. 
%From each temperature dependence illustrated as black, orange, and red dashed curves in the inset of Fig.~\ref{HC}(a), 
Then, the temperature dependence of a magnetic entropy is estimated as black open circles in Fig.~\ref{HC}(d).
A black solid curve corresponds to the magnetic entropy of a $S$ =~5/2 pyrochlore antiferromagnet with $J$ = 2.0 $\pm$ 0.2~K.
It is estimated by integrating $C_\mathrm{mag}/T$ derived from the [4, 4] Pad\'e approximant between the temperature $T$ and 2000~K,
and then subtracting the integrated value from a total entropy of $R$ ln 6.
Because of the high consistency in the estimatation of $J$ from magnetization and heat capacity measurements,
% with those of a $S$ =~5/2 pyrochlore antiferromagnet,
the magnetic entropy should also follow that expected for a $S$ =~5/2 pyrochlore antiferromagnet.
However, the magnetic entropy estimated from the above assumption (red circles) is much smaller than that expected for the pyrochlore antiferromagnet (black solid curve).
An offset should be added to the temperature dependence to compensate the underestimate of the magnetic entropy below 0.5~K, as shown by the red dashed curve.
To achieve a good agreement, the magnetic entropy of more than 4~J~K$^{-2}$~mol$^{-1}$, which corresponds to one-fourth of the total entropy, should persist at 0.5~K.
%Among the three assumptions (1), (2), and (3), the red dashed curve from (3) best agrees with the black curve.
A large degeneracy should be present near the ground state, as expected for classical pyrochlore antiferromagnets.
\textcolor{black}{Note that heat capacity from nuclear spins, which can exhibit a strong Schottky anomaly, is neglected in the above discussion.
The deficiency of the magnetic entropy above 0.5 K can be even larger if contributions from nuclear spins are not negligible.}

\section{\label{NPDexp}Neutron diffraction experiments}
\begin{figure}[t]
\centering
\includegraphics[width=8cm]{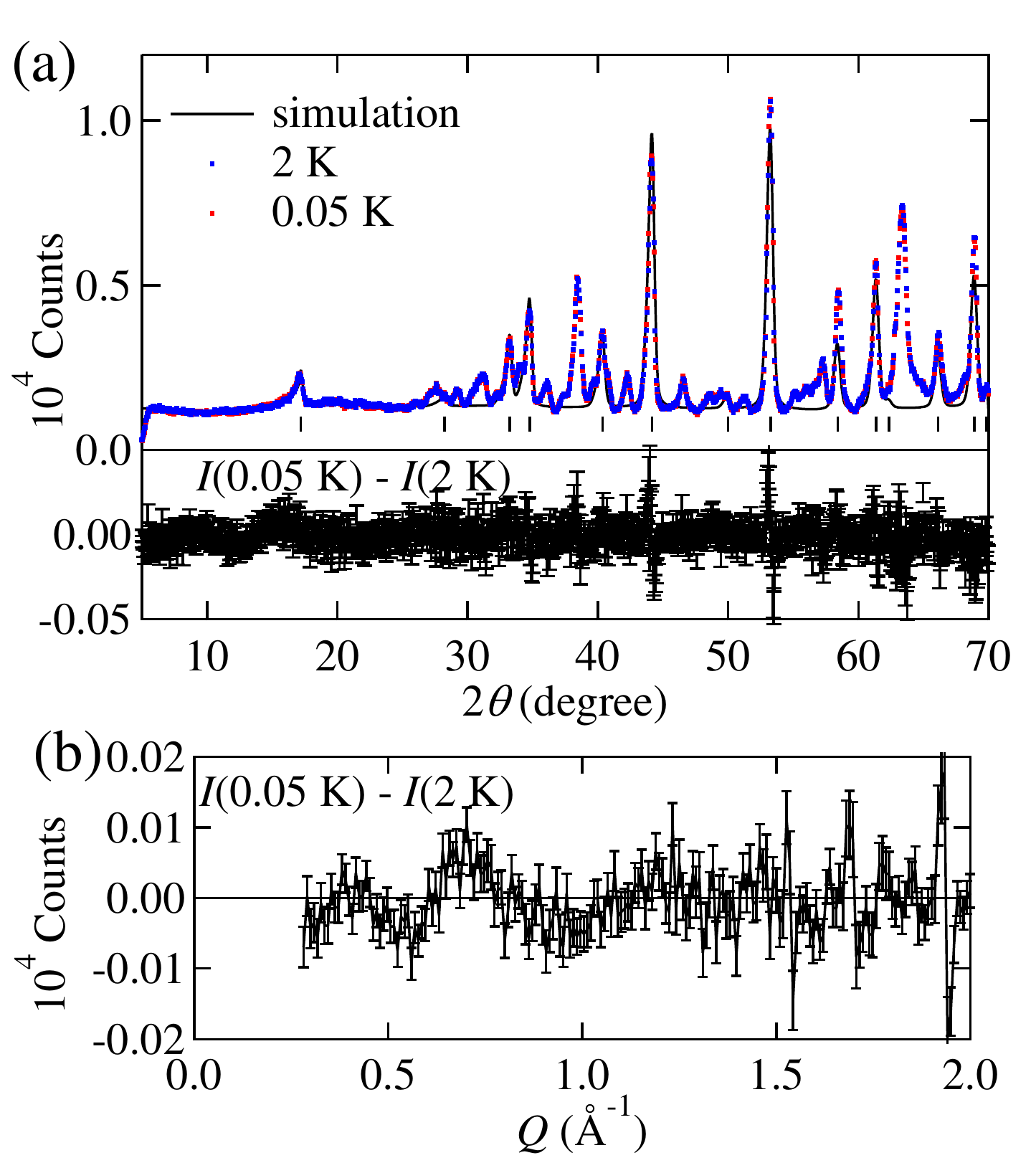}
\caption{\label{dil} (a) Neutron powder diffraction patterns measured at 0.05~K (red) and 2~K (blue) compared with simulated patterns (black). Total neutron counts are normalized by $6.0 \times 10^5$ ncu. The vertical black lines indicate the positions of the nuclear reflections. The lower panel indicate the intensity difference between the data at 0.05 K and 2 K.
\textcolor{black}{(b) Intensity difference between the data at 0.05 K and 2 K as a function of $Q$.
Data points are binned to improve statistics.}}
\label{LTpattern}
\end{figure}

The magnetization and heat capacity measurements reveal that magnetic properties of Na$_3$Mn(CO$_3$)$_2$Cl are well described by a $S$ =~5/2 nearly Heisenberg pyrochlore antiferromagnet.
In this section, we present results of the neutron diffraction experiments which evidence that a short range order characteristic of a pyrochlore antiferromagnet develops below 50~K,
while an apparent signature of the magnetic order is absent down to 0.05~K.
Figure~\ref{NPD} represents neutron diffraction patterns measured at 1.5 and 200~K.
As we discussed in section~\ref{CS}, the neutron diffraction pattern at 1.5~K is well fitted by nuclear reflections.
Only very weak unindexed peaks are left at 26.9$^\circ$ and 35.6$^\circ$, which are likely from a small amount of impurities.
While a large change in the peak intensity is often expected at a small 2$\theta$ angle if a magnetic order occurs in a $S$ =~5/2 system, such a large change was not observed.
This is consistent with the absence of any anomalies in the magnetization and heat-capacity measurements from the main phase.
Atomic positions at 1.5 and 200~K determined from the Rietveld refinement are almost identical, supporting the validity of the refinements.
To investigate whether a magnetic order is present or not at much lower temperatures, a neutron diffraction pattern is collected at 0.05~K and compared with a 2~K pattern taken under the same condition as shown in Fig.~\ref{dil}(a).
In spite of the presence of additional Bragg peaks from $d$-isopropanol, which is used as heat conduction medium in the sample can,
diffraction patterns at 0.05 and 2~K do not show \textcolor{black}{any additional peaks} up to 2$\theta$ =~70$^\circ$.
Note that Na$_3$Co(CO$_3$)$_2$Cl exhibits an all-in-all-out order at 1.5~K, evidenced by heat capacity and neutron diffraction experiments\cite{PyCo2}.
Even if the all-in-all-out order is present, the upper limit on the moment size would be 0.5~$\mu_\mathrm{B}$ according to the experimental error around 022 reflections.
This is only 10~\% of 5~$\mu_\mathrm{B}$ expected for Mn$^{2+}$ ions, and unreasonably small for a nearly-classical system. 
This clearly indicates that the magnetic order should be largely suppressed owing to magnetic frustration in Na$_3$Mn(CO$_3$)$_2$Cl.

\begin{figure}[t]
\centering
\includegraphics[width=8cm]{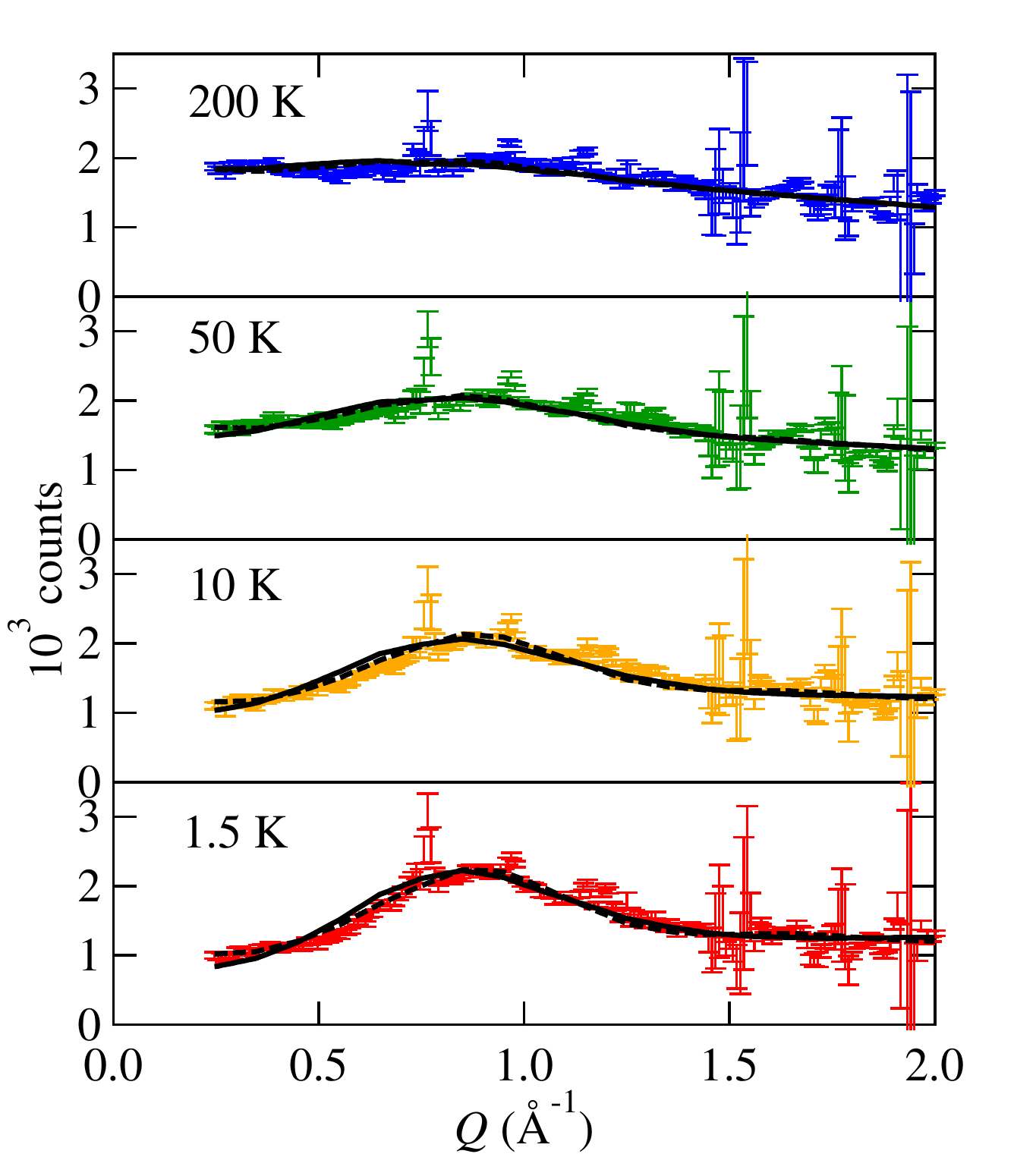}
\caption{\label{diffuse} $Q$ dependence of the intensity measured at 1.5, 10, 50, and 200~K.
\textcolor{black}{Data points are binned to improve statistics.}
Total counts are normalized by 1.5 $\times$ 10$^6$ ncu.
A solid and dashed curves represent a fit for a pyrochlore antiferromagnet with NN interactions ($J =~2$~K),
and that for a pyrochlore antiferromagnet with NN and NNN interactions ($J =~2$~K, $J' =~-0.4$~K), respectively.}
\end{figure}

Although no strong magnetic Bragg peaks are found, a small change in the intensity is observed at low scattering angles.
Figure~\ref{diffuse} represents the intensities at low scattering angles plotted as a function of a scattering wave vector $Q$. 
The background determined from empty can measurements, and intensities of nuclear reflections estimated by the Rietveld refinement (Fig.~\ref{NPD}) are subtracted.
As temperature is decreased, a broad peak develops at 0.85~\AA $^{-1}$, indicating the development of an antiferromagnetic short range order.
\textcolor{black}{The development persists down to very low temperatures, as indicated by the intensity difference shown in Fig.~\ref{dil}(b).} 
To investigate its origin, we calculated neutron-scattering cross sections using mean-field approximation\cite{Py1, MFT, MFT2, MFT3}, which is useful for simply estimating approximate exchange constants.
To avoid confusions, we use the same notations as those in Ref.~\onlinecite{MFT3}.
A pyrochlore lattice is regarded as a face-centered cubic lattice with four sublattices in each unit cell.
The positions of the $i$-th Bravais lattice point, and $a$-th sublattice in every unit cell are denoted by $\mathbf{R}_{i}$ and $\mathbf{r}^a$, respectively.
The relative position from the $b$-th sublattice in the $j$-th Bravais lattice to the $a$-th sublattice in the $i$-th Bravais lattice is defined as $\mathbf{R}^{ab}_{ij} \equiv~\mathbf{r}^{a} + \mathbf{R}_{i} - \mathbf{r}^{b} - \mathbf{R}_{j}$.
A unit vector along $\mu$ = $x$, $y$, or $z$ direction is represented by $\mathbf{n}^\mu$.  

First, Fourier transformations of nearest neighbor (NN), next-nearest neighbor (NNN), and dipolar interactions (denoted by $J$, $J'$, and $D_\mathrm{dd}$, respectively) are summed up for each sublattice and $x$, $y$, and $z$ component, resulting in a 12 $\times$ 12 matrix,
\begin{align}
J^{ab}_{\mu \nu}(\mathbf{q}) &= -J \sum_i \delta_{ab:\mathrm{NN}} (\mathbf{n}^{\mu} \cdot \mathbf{n}^{\nu}) \mathrm{exp}(-i\mathbf{q} \cdot \mathbf{R}^{ab}_{ij}) \notag \\
& \ -J' \sum_i \delta_{ab:\mathrm{NNN}} (\mathbf{n}^{\mu} \cdot \mathbf{n}^{\nu}) \mathrm{exp}(-i\mathbf{q} \cdot \mathbf{R}^{ab}_{ij}) \label{JFT} \\
& \ +D_\mathrm{dd} \sum_i \left( \frac{\mathbf{n}^{\mu} \cdot \mathbf{n}^{\nu}}{|\mathbf{R}^{ab}_{ij}|^3} - \frac{ 3 (\mathbf{n}^{\mu} \cdot \mathbf{R}^{ab}_{ij}) (\mathbf{n}^{\nu} \cdot \mathbf{R}^{ab}_{ij})} {|\mathbf{R}^{ab}_{ij}|^5} \right) \notag \\
&\ \ \ \times R_\mathrm{NN}^3  \mathrm{exp}(-i\mathbf{q} \cdot \mathbf{R}^{ab}_{ij}), \notag
\end{align}
which is defined for every wavevector $\mathbf{q}$ in the first Brillouin zone. 
$R_\mathrm{NN}$ is a distance of the nearest neighbor interactions.
Note that the signs of $J$, $J'$, and $D_\mathrm{dd}$ are opposite from those in Ref.~\onlinecite{MFT3} since we defined an antiferromagnetic exchange to be positive. 
Kronecker delta is included in eq.~\eqref{JFT} to take a sum only between nearest or next-nearest neighbors.
The lattice sum of dipolar interactions is treated via the Ewald summation method\cite{MFT3, Evaldsum, Evaldsum2, Evaldsum3}.

From $\lambda^\alpha_\mu(\mathbf{q})$ and eigenstates $\mathbf{U}^\alpha_\mu(\mathbf{q})$ ($\alpha$ =~1, 2, 3, 4 and $\mu$ =~$x^\prime$, $y^\prime$, $z^\prime$ represent normal modes) of the matrix, the mean-field neutron-scattering cross section can be estimated by,  
\begin{equation}
\begin{split}
\frac{d\sigma(\mathbf{Q})}{d\Omega} &= \frac{C}{3} \{ f(Q) \}^2 \sum_{\alpha, \mu} \frac{|\mathbf{F}^\alpha_{\mu \perp}(\mathbf{q})|^2}{1 - \lambda^\alpha_\mu(\mathbf{q})/(3 T)} \\
\mathbf{F}^{\alpha}_{\mu \perp}(\mathbf{q}) &\equiv \sum_{a} \{ \mathbf{U}^{\alpha, a}_{\mu} (\mathbf{q}) -  (\mathbf{U}^{\alpha, a}_{\mu} (\mathbf{q}) \cdot \mathbf{Q}) \mathbf{Q} \} \\
& \ \times \mathrm{exp}(i\mathbf{G} \cdot \mathbf{r}^a) \\
\mathbf{U}^{\alpha, a}_{\mu} (\mathbf{q}) &= \sum_u \mathbf{n}^u U^{\alpha, a}_{\mu, u} (\mathbf{q}) = \sum_u \mathbf{n}^u U^{*a, \alpha}_{u,\mu} (\mathbf{q}),
\end{split}
\label{scatt}
\end{equation}
where $U^{a, \alpha}_{u, \mu}$ is an $(a, u)$-component of $\mathbf{U}^\alpha_\mu$,
$\mathbf{Q}$ is related with a reciprocal wavevector $\mathbf{G}$ as $\mathbf{Q} = \mathbf{q} + \mathbf{G}$,
and $C$ and $f(Q)$ denote an arbitrary coefficient and a magnetic form factor, respectively.
In the mean-field approximation, correlation functions are regarded as the sum of a zeroth-order term of self-correlations and a first-order term proportional to an inverse temperature\cite{Py1, MFT3}.
This rough approximation is not valid at low temperatures, since the mean-field neutron-scattering cross section diverges at $T_\mathrm{C}$ which is defined by the maximum of $\lambda^\alpha_\mu(\mathbf{q})/3$ among all $\alpha$, $\mu$ and $\mathbf{q}$.
This disadvantage does not matter for the following discussion on the $Q$-dependence,
though it makes the temperature variation of the cross section unreliable.
Thus, $C$ is used as an adjustable parameter which varies with the temperature for each fit.
In addition, a constant temperature-independent background which best reproduces the 1.5~K scan is also added in the fit.

%However, it should be applicable to the present system since a magnetic transition temperature, if any, is much lower than 1.3~K. 

The neutron diffraction patterns are compared with a powder average of the scattering-cross section estimated from eq.~\eqref{scatt}, as presented in Fig.~\ref{diffuse}.
A solid curve represents a fit to the $J$--$D_\mathrm{dd}$ pyrochlore model.
In the fit, $J$ is fixed to 2~K estimated from magnetization measurements, $J'$ is fixed to 0, and $D_\mathrm{dd}$ is set to $-$0.125~K estimated from dipolar interactions expected for Mn$^{2+}$ (5~$\mu_\mathrm{B}$).
This simple model with only one fitting parameter $T$ well reproduces the broad peak due to antiferromagnetic nearest-neighbor interactions $J$;
the effect of the dipolar interactions $D_\mathrm{dd}$ is small.
Antiferromagnetic $J$ yields the scattering cross section which extends between $\mathbf{Q}$ =~(0,0,2) and (1,1,2), and $\mathbf{Q}$ = (1,1,0) and (2,2,0).
%While the scattering cross section caused from the ground state has little $\mathbf{Q}$ dependence because of very small dispersion, 
%the first excited state yields the maximum intensity at $\mathbf{Q}$ =~---, where the eigenenergy becomes the smallest.
The fit is also performed by the $J$--$J'$--$D_\mathrm{dd}$ pyrochlore model, as indicated by a dashed curve in Fig.~\ref{diffuse}.
Introducing ferromagnetic $J'$ in addition to $J$ and $D_\mathrm{dd}$ slightly sharpens the peak.
This is because it induces the energy minimum of the ground state at $\mathbf{q}$ =~($h$,$h$,0) ($h \sim$~3/4)\cite{PyJ1J23, PyJ1J24}, and thus the scattering cross section at $\mathbf{Q}$ =~($h$,$h$,2) and (2$-h$,2$-h$,0) is increased.
The best fit is achieved when $J'$ is weakly ferromagnetic, $-$0.4~K.
Weakly ferromagnetic $J'$ is also suggested for Na$_3$Co(CO$_3$)$_2$Cl\cite{PyCo2}.
As shown in Fig.~\ref{diffuse}, $Q$-dependence measured at high temperatures is also well reproduced by both models.
For the $J$--$D_\mathrm{dd}$ pyrochlore model the transition temperature becomes $T_\mathrm{C}^\mathrm{MF}$ = 1.5~K according to the mean-field theory,
and the fit yields the temperature of $T/T_\mathrm{C}^\mathrm{MF}$ =~1.7, 2.3, 6.0, and 14 for the 1.5, 10, 50, and 200~K data, respectively.
On the other hand, the $J$--$J'$--$D_\mathrm{dd}$ pyrochlore model leads to $T_\mathrm{C}^\mathrm{MF}$ = 2.0~K,
and the fit yields the temperature of $T/T_\mathrm{C}^\mathrm{MF}$ =~1.6, 2.2, 4.5, and 10 for the 1.5, 10, 50, and 200~K data, respectively.
It is difficult to conclude whether or not ferromagnetic $J'$ is present or not only from the two fits because of their small difference.
%Investigating diffuse scattering using single crystal is necessary for this purpose.
However, it is more likely that next-nearest-neighbor interactions $J'$ are very small, since a magnetic order is likely to be absent above 0.05~K.
A transition temperature to the multi-$Q$ order increases with increasing next-nearest neighbor interactions $J'$
according to the Monte Carlo simulations\cite{PyJ1J23, PyJ1J24}. 
If $J'$ of $-$0.4~K is present, multi-$Q$ order should appear near 2~K.

As discussed in previous and present sections, magnetic properties of Na$_3$Mn(CO$_3$)$_2$Cl are consistent with those of a classical pyrochlore antiferromagnet.
A temperature dependence of the magnetic susceptibility and heat capacity, the magnetization curve, $Q$-dependence of the diffuse scattering are well explained by those of the pyrochlore antiferromagnet.
In addition, magnetization, heat capacity and neutron diffraction patterns exhibit no apparent signature of a spin-glass transition above 2~K or a magnetic order above 0.05~K.
Furthermore, the estimation of the magnetic entropy indicates a large degeneracy near the ground state.
From these features, it is likely that Na$_3$Mn(CO$_3$)$_2$Cl is a good candidate of a classical nearly-Heisenberg pyrochlore antiferromagnet; extra interactions which lift the ground state degeneracy should be small.
A single-ion or an exchange anisotropies should be absent in Na$_3$Mn(CO$_3$)$_2$Cl, while they may lead to the all-in-out-order in Na$_3$Co(CO$_3$)$_2$Cl\cite{PyCo2}.
%This may be due to the structure consists of MnO$_6$ octahedra separated by carbonate ions.
In addition, small dipole interactions cannot lift the line degeneracy along (111) directions\cite{Gd2Ti2O7_1}.
It should be noted that the possibility of a spin-glass transition below 2~K is small from the thorough consistency.
To reveal spin dynamics in this compound, low energy excitations should be further examined by neutron scattering, NMR or $\mu$SR experiments.
%This feature is in contrast with the all-in-all-out order observed in Na$_3$Co(CO$_3$)$_2$Cl.
%This is because a single-ion or an exchange anisotropies are absent in Na$_3$Mn(CO$_3$)$_2$Cl, while they can be present in Na$_3$Co(CO$_3$)$_2$Cl because of an orbital degeneracy.
%In addition, note that a disordered ground state of Na$_3$Mn(CO$_3$)$_2$Cl is quite different from the all-in-out-order ground state found in Na$_3$Mn(CO$_3$)$_2$Cl\cite{PyCo2}.

\section{\label{summary}Summary}
Magnetic properties of Na$_3$Mn(CO$_3$)$_2$Cl are investigated by means of magnetization, heat capacity, and neutron diffraction experiments. 
The temperature dependence of the magnetic susceptibility and the magnetization curve are consistent with those of a $S$ =~5/2 pyrochlore lattice antiferromagnet with nearest-neighbor interactions.
The neutron diffraction patterns indicate development of the antiferromagnetic short-range order below 50~K.
No apparent signature of a spin-glass transition and a magnetic order is observed above 2 and 0.05~K, respectively.
A large degeneracy should remain near the ground state, as indicated the heat capacity measurement;
the magnetic entropy of 4 J K$^{-2}$ mol$^{-1}$ still remains at 0.5 K.
In this sense, Na$_3$Mn(CO$_3$)$_2$Cl should exhibit a interesting disordered ground state as expected for a classical pyrochlore antiferromagnet.

\begin{acknowledgments}
We thank H. Sagayama for supporting a single crystal XRD experiment,
\textcolor{black}{S. Asai and T. Masuda for a help to measure heat capacity,}
and T. Hiraiwa and K. Yamauchi for fruitful discussions.
This work was partly supported by Grants-In-Aid for Scientific Research (Grants No. 17K18744) from MEXT of Japan,
the Research Program for CORE lab of ''Dynamic Alliance for Open Innovation Bridging Human, Environment and Materials'' in ''Network Joint Research Center for Materials and Devices'',
and the Motizuki Fund of Yukawa Memorial Foundation.
The high-field magnetization and the heat capacity measurements were carried out under the Inter-University Cooperative Research Program of the Institute for Materials Research, Tohoku University,
and the Visiting Researcher's Program of the Institute for Solid State Physics, the University of Tokyo, respectively. 
The synchrotron x-ray diffraction experiments were performed with the approval of the Photon Factory Program Advisory Committee (No.2016G143).
Travel expenses for the experiment on ECHIDNA at ANSTO were partly sponsored by the General User Program of ISSP-NSL, University of Tokyo.
\end{acknowledgments}

\end{document}